\documentclass[aps,pre,
onecolumn,
groupedaddress,showpacs]{revtex4-2}

\usepackage{threeparttable}
\usepackage{epsfig}
\usepackage{epstopdf}
\usepackage{appendix}
\usepackage{graphicx}

\usepackage{amsmath}
\usepackage[T1]{fontenc}
\usepackage{caption}
\usepackage{float} 
\usepackage{dcolumn}
\usepackage{hyperref}
\usepackage{amsthm}
\usepackage{amssymb}

\usepackage{stackengine}
\newcommand\xrowht[2][0]{\addstackgap[.5\dimexpr#2\relax]{\vphantom{#1}}}
\usepackage{booktabs}
\usepackage{xparse}
\usepackage{tikz}
\usetikzlibrary{calc}

\tikzset{Arrow Style/.style={text=black, font=\boldmath}}

\newcommand*{\XShift}{0.5em}
\newcommand*{\YShift}{0.5ex}

\NewDocumentCommand{\DrawArrow}{s O{} m m m}{%
    \begin{tikzpicture}[overlay,remember picture]
        \draw[->, thick, Arrow Style, #2] 
                ($(#3.west)+(\XShift,\YShift)$) -- 
                ($(#4.east)+(-\XShift,\YShift)$)
        node [midway, above] {#5};
    \end{tikzpicture}%
}

\def\e{\mbox{e}}
\def\d{\mbox{d}}

\begin{document}


\title{Infinite series involving special functions obtained using simple one-dimensional quantum mechanical problems}
\author{Sonja Gombar$^{1}$} \email[]{sonja.gombar@df.uns.ac.rs} \author{Milica Rutonjski$^{1}$} \author{Petar Mali$^{1}$}  \author{Slobodan Rado\v sevi\' c$^{1}$} \author{Milan Panti\' c$^{1}$} \author{Milica Pavkov-Hrvojevi\' c$^{1}$}
\affiliation{$^1$ Department of Physics, Faculty of Sciences, University of Novi Sad,
Trg Dositeja Obradovi\' ca 4, 21000 Novi Sad, Serbia}

\date{\today}


\begin{abstract}
 In this paper certain classes of infinite sums involving special functions are evaluated analytically by application of basic quantum mechanical principles to simple models of half harmonic oscillator and a particle trapped inside an infinite potential well. The infinite sums $\sum^{\infty}_{n=0}\frac{2^{2n}}{(2n+1)!}\Gamma^{2}\left(n+\frac{3}{2}\right)\left[\hspace{0.2mm}_2\hspace{-0.03cm}F_1\left(-n,\frac{\nu+2}{2};\frac{3}{2};\frac{1}{2}\right)\right]^{2}$, $\sum^{\infty}_{n=0}\frac{\left[L_{\nu}^{2n+1-\nu}\left(\frac{b^{2}}{2}\right)\right]^{2}b^{4n}}{2^{2n}(2n+1)!}$ and $\sum^{\infty}_{n=1}\frac{\big[J_{\nu+1}(n\pi)\big]^{2}}{n^{2\nu}}$, where $_2\hspace{-0.03cm}F_1\left(-n,\frac{\nu+2}{2};\frac{3}{2};\frac{1}{2}\right)$ is generalized hypergeometric function, $L_{\nu}^{2n+1-\nu}\left(\frac{b^{2}}{2}\right)$ associated Laguerre polynomial and $J_{\nu+1}(n\pi)$ Bessel function of the first kind, are calculated for integer $\nu$. It is also demonstrated that the same procedure can be generalized by application to some classes of functions which are not regular wave functions leading to additional infinite sums, as a consequence of which the series $\sum_{n=1}^{\infty}\frac{\left[\mathsf{H}_{\nu}(n\pi)\right]^{2}}{n^{2\nu}}$ containing Struve functions of the first kind $\mathsf{H}_{\nu}(n\pi)$ are evaluated. Convergence of the evaluated series, additionally verified by the application of different convergence tests, is secured by the properties of the corresponding Hilbert space.

\end{abstract}
\maketitle

\section{Introduction}
\label{intro}

Although the roots of physical mathematics were planted back in ancient Greece \cite{levi,uspeh}, its present form and significance were shaped in George Polya's work \cite{polya}. Following in his footsteps in an effort to provide novel mathematical results, building upon various physical problems, this paper serves as yet another illustration of the unbreakable bond between physics and mathematics. Furthermore, the present paper aims to show that even slightly altered physical solutions offer a productive framework for obtaining additional mathematical results.

In physical problems, one often comes across different mathematical expressions containing special functions \cite{nikiuva,rad1,rad2,rad3,rad4,rad5}. Namely, various studies indicate that the evaluation of some infinite sums involving special functions represents a practical problem \cite{rad6,rad7,rad8,rad9}.  Our recent work \cite{mali,mali2} has shown that by making use of quantum mechanical principles some previously unevaluated infinite sums can be obtained analytically. Specifically, using a simple quantum mechanical model of a particle in a box, we were able to calculate several classes of infinite sums with some special functions such as Bessel functions of the first kind and generalized hypergeometric functions. 

In the current paper we further exploit the model of a particle trapped inside an infinite potential well and also extend our research to the model of a half harmonic oscillator. Obtained infinite sums include several types of special functions. The first type involves generalized hypergeometric functions ${_p\hspace{-0.03cm}F_{q}}(\alpha_1,\alpha_2,...,\alpha_p;\beta_1,\beta_2,...,\beta_{q};z)$ defined by the series \cite{GRRI} 
\begin{equation}
{_p\hspace{-0.03cm}F_{q}}(\alpha_1,\alpha_2,...,\alpha_p;\beta_1,\beta_2,...,\beta_{q};z
)=  \sum_{l=0}^{\infty}\frac{(\alpha_{1})_{l}(\alpha_{2})_{l}...(\alpha_{p})_{l}\,z^{l}}{(\beta_{1})_{l}(\beta_{2})_{l}...(\beta_{q})_{l}\,l!}  
\end{equation}
if the series converges. Here $p$ and $q$ are integers and $(\alpha)_{l}=\frac{\Gamma(\alpha+l)}{\Gamma(\alpha)}$ are Pochhammer symbols, defined
in terms of Euler's gamma functions $\Gamma(\alpha)$. The aforementioned series converges absolutely for all $z$ if $p<q+1$, whereas it diverges for all $z\neq 0$ if $p>q+1$. If $p=q+1$, the series converges absolutely for $|z|<1$. The series also converges absolutely for $|z|=1$ if $\sum_{j=1}^{q}\beta_{j}>\sum_{i=1}^{p}\alpha_{i}$.
The required conditions for $\beta_{j}$ ($j=1,2,3,...,q$) are that $\beta_{j}\neq 0$ and $\beta_{j}\notin \mathrm{Z}_{-}$ \cite{sloba}. Specifically, for $p=q=1$ generalized hypergeometric function is known as Kummer's hypergeometric function, whereas for $p=2$ and $q=1$ it is known as Gaussian hypergeometric function. The next class of sums involves associated Laguerre polynomials $L_{p}^{q}(z)$ given by \cite{GRRI}
\begin{equation}
L_{p}^{q}(z)=\frac{1}{p!}\e^{z}z^{-q}\frac{\d^{p}}{\d z^{p}}\left(\e^{-z}z^{p+q}\right),   
\end{equation}
where $p,q=0,1,2,3,...$ The third type of infinite sums includes the Bessel functions of the first kind $J_p(z)$ defined by series \cite{GRRI}
\begin{equation}
J_p(z)=\left(\frac{z}{2}\right)^{p}\sum_{l=0}^{\infty}\frac{(-1)^{l}}{l!\,\Gamma(p+l+1)}\left(\frac{z}{2}\right)^{2l}, \hspace{2mm} |\mathrm{arg}z|<\pi.    
\end{equation}
The last type of special functions occurring in the evaluated infinite sums is the Struve function of the first kind $\mathsf{H}_p(z)$ with the power series form \cite{GRRI}
\begin{equation}
\mathsf{H}_{p}(z)=\sum_{l=0}^{\infty}(-1)^{l}\frac{\left(\frac{z}{2}\right)^{2l+p+1}}{\Gamma\left(l+\frac{3}{2}\right)\Gamma\left(p+l+\frac{3}{2}\right)}.    
\end{equation}

The first of the discussed models is a simple model of a half harmonic oscillator, with corresponding Hamiltonian $\hat{\mathcal{H}}=-\frac{\hbar^2}{2m}\frac{\mathrm{d}^2}{\mathrm{d} x^2}+V(x)$, where $m$ denotes particle's mass, $V(x)$ presents the potential defined by
\begin{equation}
    V(x)=\begin{cases} \infty, & x\leq 0 \\
    \frac{1}{2}m\omega^{2}x^{2}, & x>0
    \end{cases}
\end{equation}
and $\omega$ is the angular frequency of the classical oscillator. The normalized eigenstates and eigenvalues of the aforementioned Hamiltonian are
\begin{equation}
    \psi_{n}(x)=
    \sqrt{\frac{\sqrt{\alpha}}{2^{2n}(2n+1)!\sqrt{\pi}}}H_{2n+1}(\sqrt{\alpha}x)\e^{-\frac{\alpha x^{2}}{2}},
    \hspace{5mm} \alpha=\frac{m\omega}{\hbar},
\end{equation}
defined on the interval $x>0$, and
\begin{equation}
    E_{n}=\left(2n+\frac{3}{2}\right)\hbar\omega, \hspace{2mm} n=0,1,2,3,...
\end{equation}
respectively. Hermite polynomials $H_{n}(z)$ are given by \cite{GRRI}
\begin{equation}
H_{n}(z)=(-1)^{n}\e^{z^{2}}\frac{\d^{n}}{\d z^{n}}\left(\e^{-z^{2}}\right).
\end{equation}
Without loss of generality we set $\alpha=1$, which leads to 
\begin{equation}
    \psi_{n}(x)=
    \sqrt{\frac{1}{2^{2n}(2n+1)!\sqrt{\pi}}}H_{2n+1}(x)\e^{-\frac{x^{2}}{2}}.
\end{equation}
From the superposition principle 
\begin{equation}
    \psi(x)=\sum_{n=0}^{\infty}C_{n}\psi_{n}(x), \label{lhokoef}
\end{equation}
where $\psi(x)$ denotes an arbitrary state of a half harmonic oscillator, one can calculate expansion coefficients using
\begin{equation}
    C_{n}=\int_{0}^{\infty}\psi_{n}^{*}(x)\psi(x)\d x. \label{lhokoef2}
\end{equation}    

The second exploited model, one of a particle of mass $m$ trapped inside an infinite potential well
\begin{equation}
    V(x)=\begin{cases} 0, & x\in(0,a) \\
    \infty, & x\notin(0,a)
    \end{cases},
\end{equation}
where $a$ is the well width, has already been discussed in \cite{mali,mali2}. In the present paper we make further use of this model to compute additional sums, which have not been previously calculated. In this case the normalized eigenstates and the corresponding eigenvalues are
\begin{equation}
    \varphi_{n}(x)=
        \sqrt{\frac{2}{a}}\sin\frac{n\pi x}{a},
\end{equation}
defined on the interval $x\in(0,a)$, and
\begin{equation}
    E_{n}=\frac{n^{2}\pi^{2}\hbar^{2}}{2ma^{2}}, \hspace{2mm} n=1,2,3,4,...
\end{equation}
respectively. Analogously to the previous case, expansion coefficients from the equation
\begin{equation}
    \varphi(x)=\sum_{n=1}^{\infty}C_{n}\varphi_{n}(x), \label{jamakoef}
\end{equation}
where $\varphi(x)$ denotes an arbitrary state of a particle trapped in an infinite potential well, can be determined from
\begin{equation}
    C_{n}=\int_{0}^{a}\varphi_{n}^{*}(x)\varphi(x)\d x. \label{jamakoef2}
\end{equation}
Without loss of generality we restrict ourselves to the case of real functions $\psi(x)$ and $\varphi(x)$.

The infinite sums calculated in the present paper are obtained using normalization condition for coefficients \eqref{lhokoef2} and \eqref{jamakoef2}. \textit{It should be noted that this method secures the convergence of the obtained infinite series, independently of the choice of the quantum mechanical problem.} 

The rest of the paper is organized as follows: The results are introduced in Sec. II, and Sec. III concludes the paper. 

\section{Results}
\label{Results}

\subsection{Half harmonic oscillator}

We start with the wave function
\begin{equation}
\psi(x)=\sqrt{2 \frac{3^{\nu+1/2}}{\Gamma \left( \nu + \frac 12  \right)}}\e^{-\frac{3x^{2}}{2}}x^{\nu},  
 \label{prvafunk}   
\end{equation}
defined on the interval $x>0$, where $\nu$ takes on positive integer values. For instance, $\nu=1$ corresponds to the wave function
\begin{equation}
 \psi(x)=\sqrt{\frac{12\sqrt{3}}{\sqrt{\pi}}}\e^{-\frac{3x^{2}}{2}}x.  \label{wf1} 
\end{equation}
In that case, coefficients in the infinite series \eqref{lhokoef} are
\begin{equation}
C_{n}=\int_{0}^{\infty}\psi_{n}^{*}(x)\psi(x)\d x=\sqrt{\frac{12\sqrt{3}}{\pi2^{2n}(2n+1)!}}(-1)^{n}2^{2n-\frac{1}{2}}\frac{\Gamma\left(\frac{3}{2}\right)\Gamma\left(n+\frac{3}{2}\right)}{\sqrt{\pi}}\hspace{0.2mm}_2\hspace{-0.03cm}F_1\left(-n,\frac{3}{2};\frac{3}{2};\frac{1}{2}\right). \label{coef}    
\end{equation}
From the condition $\sum^{\infty}_{n=0}|C_n|^2=1$, we obtain
\begin{equation}
\sum^{\infty}_{n=0}\frac{2^{2n}}{(2n+1)!}\Gamma^{2}\left(n+\frac{3}{2}\right)\left[\hspace{0.2mm}_2\hspace{-0.03cm}F_1\left(-n,\frac{3}{2};\frac{3}{2};\frac{1}{2}\right)\right]^{2}=\sum_{n=0}^{\infty}\frac{\Gamma^{2}\left(n+\frac{3}{2}\right)}{(2n+1)!}=\frac{2\pi}{3\sqrt{3}}, 
\end{equation}
where we used \cite{GRRI}
\begin{equation}
\hspace{0.2mm}_2\hspace{-0.03cm}F_1\left(-n,\frac{3}{2};\frac{3}{2};\frac{1}{2}\right)=\left(1-\frac{1}{2}\right)^{n}=\left(\frac{1}{2}\right)^{n}. \label{hyper3/2}
\end{equation}
This result, which corroborates the aforementioned statement that for a well-defined quantum mechanical problem the series convergence is secured by the corresponding Hilbert space properties, is additionally verified by d'Alembert's ratio convergence test (for more details see App. II). In \eqref{coef} we used \cite{GRRI,erdel}
\begin{equation}
\int_{0}^{\infty}\e^{-2 x^{2}}x^{\nu}H_{2n+1}(x)\d x=(-1)^{n}2^{2n-\frac{1}{2}\nu}\frac{\Gamma\left(\frac{\nu}{2}+1\right)\Gamma\left(n+\frac{3}{2}\right)}{\sqrt{\pi}}\hspace{0.2mm}_2\hspace{-0.03cm}F_1\left(-n,\frac{\nu}{2}+1;\frac{3}{2};\frac{1}{2}\right),\hspace{2mm} \mathrm{Re}\hspace{1mm}\nu>-2.  \label{hipergeo}   
\end{equation}
Following this procedure leads to a class of infinite sums labeled by $\nu$, $\sum^{\infty}_{n=0}\frac{2^{2n}}{(2n+1)!}\Gamma^{2}\left(n+\frac{3}{2}\right)\left[\hspace{0.2mm}_2\hspace{-0.03cm}F_1\left(-n,\frac{\nu+2}{2};\frac{3}{2};\frac{1}{2}\right)\right]^{2}$, some of which can be found in Table \ref{table1}. Since the gamma function can be calculated in an exact way only for integer and half-integer arguments, for the wave functions of type \eqref{prvafunk} we considered only integer values of $\nu$. 

\begin{widetext}
\begin{center}
\begin{table}
\caption{List of the series $\sum^{\infty}_{n=0}\frac{2^{2n}}{(2n+1)!}\Gamma^{2}\left(n+\frac{3}{2}\right)\left[\hspace{0.2mm}_2\hspace{-0.03cm}F_1\left(-n,\frac{\nu+2}{2};\frac{3}{2};\frac{1}{2}\right)\right]^{2}=\frac{\pi^{3/2} 2^{\nu-1} \Gamma \left(\nu + \frac{1}{2}  \right)}{3^{\nu+1/2} \left[  \Gamma \left(   \frac{\nu}{2} +1  \right)  \right]^2}
$.}
\begin{threeparttable}
 \begin{tabular}{|c|c|c| } 

 \hline\hline
  $\nu$ & $\psi(x)$ & sum of the infinite series \\ [0.5ex] 
 \hline\hline \xrowht[()]{10pt}
 $1$ & $\sqrt{\frac{12\sqrt{3}}{\sqrt{\pi}}}\e^{-\frac{3x^{2}}{2}}x$ &  $\frac{2\pi}{3\sqrt{3}}$  \\  \hline \xrowht[()]{10pt}
 
 $2$  &  $\sqrt{\frac{24\sqrt{3}}{\sqrt{\pi}}}\e^{-\frac{3x^{2}}{2}}x^{2}$ &  $\frac{\pi^{2}}{6\sqrt{3}}$   \\ \hline \xrowht[()]{10pt}

 $3$  & $\sqrt{\frac{144\sqrt{3}}{5\sqrt{\pi}}}\e^{-\frac{3x^{2}}{2}}x^{3}$ &  $\frac{40\pi}{81\sqrt{3}}$   \\ \hline \xrowht[()]{10pt}

 $4$  & $\sqrt{\frac{864\sqrt{3}}{35\sqrt{\pi}}}\e^{-\frac{3x^{2}}{2}}x^{4}$ & $\frac{35\pi^{2}}{216\sqrt{3}}$   \\ \hline \xrowht[()]{10pt}

$5$  & $\sqrt{\frac{576\sqrt{3}}{35\sqrt{\pi}}}\e^{-\frac{3x^{2}}{2}}x^{5}$ &  $\frac{224\pi}{405\sqrt{3}}$   \\ \hline \xrowht[()]{10pt}

 $6$  & $\sqrt{\frac{3456\sqrt{3}}{385\sqrt{\pi}}}\e^{-\frac{3x^{2}}{2}}x^{6}$ &  $\frac{385\pi^{2}}{1944\sqrt{3}}$   \\ \hline \xrowht[()]{10pt}

  $7$  & $\sqrt{\frac{20736\sqrt{3}}{5005\sqrt{\pi}}}\e^{-\frac{3x^{2}}{2}}x^{7}$ &  $\frac{18304\pi}{25515\sqrt{3}}$   \\ \hline \xrowht[()]{10pt}

 $8$  & $\sqrt{\frac{41472\sqrt{3}}{25025\sqrt{\pi}}}\e^{-\frac{3x^{2}}{2}}x^{8}$ &  $\frac{25025\pi^{2}}{93312\sqrt{3}}$   \\ \hline \xrowht[()]{10pt}

   $9$  & $\sqrt{\frac{248832\sqrt{3}}{425425\sqrt{\pi}}}\e^{-\frac{3x^{2}}{2}}x^{9}$ &  $\frac{1244672\pi}{1240029\sqrt{3}}$   \\ \hline \xrowht[()]{10pt}

   $10$  &  $\sqrt{\frac{1492992\sqrt{3}}{8083075\sqrt{\pi}}}\e^{-\frac{3x^{2}}{2}}x^{10}$ &  $\frac{323323\pi^{2}}{839808\sqrt{3}}$   \\ \hline

 \hline\hline 
\end{tabular}

\end{threeparttable}
\label{table1}
\end{table}
\end{center}
\end{widetext}

Also, note that for $\nu=0$ the function $\psi(x)$ must be defined as
\begin{equation}
\psi(x)=\sqrt{\frac{2\sqrt{3}}{\sqrt{\pi}}}\e^{-\frac{3x^{2}}{2}}, \hspace{2mm} x>0 \hspace{1cm} {\rm{and}} \hspace{1cm}  \psi(x)=0, \hspace{2mm} x\leq0.  \label{funkcgr}   
\end{equation}
It does not represent a regular wave function since it is not continuous at the point $x=0$. Nevertheless, one can also obtain the corresponding sum:
\begin{equation}
\boxed{\sum^{\infty}_{n=0}\frac{2^{2n}}{(2n+1)!}\Gamma^{2}\left(n+\frac{3}{2}\right)\left[\hspace{0.2mm}_2\hspace{-0.03cm}F_1\left(-n,1;\frac{3}{2};\frac{1}{2}\right)\right]^{2}=\frac{\pi^{2}}{2\sqrt{3}}} \label{diskon}   
\end{equation}
from condition $\sum^{\infty}_{n=0}|C_n|^2=1$ imposed on superposition coefficients \eqref{lhokoef2}. Functions \eqref{wf1} and \eqref{funkcgr} are shown in Figure \ref{fig:1} a) and b), respectively.

We proceed with wave function
\begin{equation}
\psi(x)=\sqrt{\frac{4\e^{b^{2}}}{\sqrt{\pi}(\e^{b^{2}}-1)}}\e^{-\frac{x^{2}}{2}}H_{0}(x)\sin(b x),  
\end{equation}
defined on the interval $x>0$, which gives coefficients \eqref{lhokoef2}
\begin{equation}
C_{n}=\sqrt{\frac{4\e^{b^{2}}}{\pi(\e^{b^{2}}-1)2^{2n}(2n+1)!}}\frac{(-1)^{n}}{2}\sqrt{\pi}b^{2n+1}\e^{-\frac{b^{2}}{4}}. \label{coef2}   
\end{equation}
Here we used \cite{GRRI,erdel}
\begin{equation}
\int_{0}^{\infty}\e^{-x^{2}}H_{0}(x)H_{2n+1}(x)\sin(bx)\d x=\frac{(-1)^{n}}{2}\sqrt{\pi}b^{2n+1}\e^{-\frac{b^{2}}{4}}L_{0}^{2n+1}\left(\frac{b^{2}}{2}\right), \hspace{2mm} b>0.\label{sinus} 
\end{equation}
The normalization condition leads to the following infinite sum:
\begin{equation}
\sum_{n=0}^{\infty}\frac{\left[L_{0}^{2n+1}\left(\frac{b^{2}}{2}\right)\right]^{2}b^{4n}}{2^{2n}(2n+1)!}=\frac{\sinh(\frac{b^{2}}{2})}{\frac{b^{2}}{2}}. \label{lag1}  
\end{equation}

Further we move on to the wave function
\begin{equation}
\psi(x)=\sqrt{\frac{2\e^{b^{2}}}{\sqrt{\pi}(1-2b^{2}+\e^{b^{2}})}}\e^{-\frac{x^{2}}{2}}H_{1}(x)\cos(b x),  
\end{equation}
defined on the interval $x>0$. Making use of \cite{GRRI,erdel}
\begin{equation}
\int_{0}^{\infty}\e^{-x^{2}}H_{1}(x)H_{2n+1}(x)\cos(bx)\d x=(-1)^{n}\sqrt{\pi}b^{2n}\e^{-\frac{b^{2}}{4}}L_{1}^{2n}\left(\frac{b^{2}}{2}\right), \hspace{2mm} b>0, \label{kosinus}
\end{equation}
we obtain coefficients
\begin{equation}
C_{n}=\sqrt{\frac{2\e^{b^{2}}}{(1-2b^{2}+\e^{b^{2}}) 2^{2n}(2n+1)!}}(-1)^{n}b^{2n}\e^{-\frac{b^{2}}{4}}L_{1}^{2n}\left(\frac{b^{2}}{2}\right),    
\end{equation}
leading to the infinite sum
\begin{equation}
\sum_{n=0}^{\infty}\frac{\left[L_{1}^{2n}(\frac{b^{2}}{2})\right]^{2}b^{4n}}{2^{2n}(2n+1)!}=\frac{1-2b^{2}+\e^{b^{2}}}{2\e^{\frac{b^{2}}{2}}}.  \label{lag2}     
\end{equation}
For more details on the convergence of the infinite sums \eqref{lag1} and \eqref{lag2} confirmed by the ratio test, see App. II. Continuing in the same manner, using functions of types $\psi(x)\sim \e^{-\frac{x^{2}}{2}}H_{\nu}(x)\sin(bx)$ for $\nu=2m$ ($m=1,2,3,...$) and $\psi(x)\sim \e^{-\frac{x^{2}}{2}}H_{\nu}(x)\cos(bx)$ for $\nu=2m+1$ ($m=1,2,3,...$) we were able to evaluate sums $\sum^{\infty}_{n=0}\frac{\left[L_{\nu}^{2n+1-\nu}\left(\frac{b^{2}}{2}\right)\right]^{2}b^{4n}}{2^{2n}(2n+1)!}$ by exploiting integrals \eqref{sin} and \eqref{cos} quoted in App. I. The examples of these sums are given in Table \ref{table4}. For more details on the calculation of integrals \eqref{hipergeo}, \eqref{sinus} and \eqref{kosinus}, see App. I.

\begin{widetext}
\begin{center}
\begin{table}
\caption{List of the series $\sum^{\infty}_{n=0}\frac{\left[L_{\nu}^{2n+1-\nu}\left(\frac{b^{2}}{2}\right)\right]^{2}b^{4n}}{2^{2n}(2n+1)!}=\frac{b^{2 (\nu-1)}}{2^{\nu} \nu!} {\rm e}^{-\frac{b^2}{2}}
    \Big{[} {\rm e}^{b^2} - (-1)^\nu L_{\nu} \left( 2 b^2  \right) \Big{]}$.}
\begin{threeparttable}
 \begin{tabular}{|c|c| } 

 \hline\hline
  $\nu$ & sum of the infinite series \\ [0.5ex] 
 \hline\hline \xrowht[()]{10pt}
 $0$ &  $\frac{2\sinh(\frac{b^{2}}{2})}{b^2}$  \\  \hline \xrowht[()]{10pt}
 
 $1$  &  $\frac{1-2b^{2}+\e^{b^{2}}}{2}\e^{-\frac{b^{2}}{2}}$   \\ \hline \xrowht[()]{10pt}

 $2$   &  $\frac{b^{2}(-1+4b^{2}-2b^{4}+\e^{b^{2}})}{8}\e^{-\frac{b^{2}}{2}}$   \\ \hline \xrowht[()]{10pt}

 $3$  & $\frac{b^{4}(3-18b^{2}+18b^{4}-4b^{6}+3\e^{b^{2}})}{144}\e^{-\frac{b^{2}}{2}}$   \\ \hline \xrowht[()]{10pt}

$4$  &  $\frac{b^{6}\big(-2b^2 (-2 + b^2) (6 - 6 b^2 + b^4) + 3 (-1 + \e^{b^{2}})\big)}{1152}\e^{-\frac{b^{2}}{2}}$   \\ \hline \xrowht[()]{10pt}

 $5$  &  $\frac{b^{8}\big(-2 b^2 (5 - 5 b^2 + b^4) (15 - 15 b^2 + 2 b^4) + 15 (1 + \e^{b^{2}})\big)}{57600}\e^{-\frac{b^{2}}{2}}$   \\ \hline \xrowht[()]{10pt}

  $6$  &  $\frac{b^{10}\big(-2 b^2 (-3 + b^2) (90 - 195 b^2 + 135 b^4 - 30 b^6 + 2 b^8) + 
  45 (-1 + \e^{b^{2}})\big)}{2073600}\e^{-\frac{b^{2}}{2}}$   \\ \hline \xrowht[()]{10pt}

 $7$ &  $\frac{b^{12}\big(-2 b^2 (2205 - 6615 b^2 + 7350 b^4 - 3675 b^6 + 882 b^8 - 98 b^{10} + 
     4 b^{12}) + 315 (1 +\e^{b^{2}})\big)}{203212800}\e^{-\frac{b^{2}}{2}}$   \\ \hline \xrowht[()]{10pt}

   $8$  &  $\frac{b^{14}\big(-2 b^2 (-2520 + 8820 b^2 - 11760 b^4 + 7350 b^6 - 2352 b^8 + 
     392 b^{10} - 32 b^{12} + b^{14}) + 315 (-1 + \e^{b^{2}})\big)}{3251404800}\e^{-\frac{b^{2}}{2}}$   \\ \hline \xrowht[()]{10pt}

   $9$ &  $\frac{b^{16}\big (-2 b^2 (-3 + b^2) (-8505 + 31185 b^2 - 42525 b^4 + 25515 b^6 - 
     7371 b^8 + 1071 b^{10} - 75 b^{12} + 2 b^{14}) + 2835 (1 + \e^{b^{2}})\big)}{526727577600}\e^{-\frac{b^{2}}{2}}$   \\ \hline

 \hline\hline 
\end{tabular}

\end{threeparttable}
\label{table4}
\end{table}
\end{center}
\end{widetext}

\subsection{Infinite potential well}

We consider first wave function of the form
\begin{equation}
\varphi(x)=\sqrt{\frac{4}{\sqrt{\pi}}\frac{1}{a^{4\nu+1}}\frac{\Gamma\left(2\nu+\frac{3}{2}\right)}{\Gamma(2\nu)}}x(a^{2}-x^{2})^{\nu-\frac{1}{2}},   
\end{equation}
defined on the interval $x\in(0,a)$. As an example, for $\nu=1$ wave function reads
\begin{equation}
\varphi(x)=\sqrt{\frac{15}{2a^{5}}}x(a^{2}-x^{2})^{\frac{1}{2}}.  \label{wf3}  
\end{equation} 
Corresponding coefficients $C_{n}$, \eqref{jamakoef2}, obtained by employing \cite{GRRI,erdel}
\begin{equation}
\int_{0}^{u}x\left(u^{2}-x^{2}\right)^{\nu-\frac{1}{2}}\sin(\tilde{a}x)\d x=\frac{\sqrt{\pi}}{2}u\left(\frac{2u}{\tilde{a}}\right)^{\nu}\Gamma\left(\nu+\frac{1}{2}\right)J_{\nu+1}(\tilde{a}u), \hspace{2mm}u>0,\hspace{2mm} \mathrm{Re}\hspace{1mm}\nu>-\frac{1}{2}  \label{besel}   
\end{equation}
are
\begin{equation}
C_{n}=\int_{0}^{a}\varphi_{n}^{*}(x)\varphi(x)\d x=\frac{\sqrt{15}}{2}\frac{J_{2}(n\pi)}{n}.    
\end{equation}
From the condition $\sum^{\infty}_{n=1}|C_n|^2=1$, we obtain
\begin{equation}
\sum_{n=1}^{\infty}\frac{\big[J_{2}(n\pi)\big]^{2}}{n^{2}}=\frac{4}{15}.    
\end{equation}
We restrict ourselves to considering integer values of $\nu$ since the half-integer ones have already been discussed in \cite{mali}. The sums $\sum^{\infty}_{n=1}\frac{\big[J_{\nu+1}(n\pi)\big]^{2}}{n^{2\nu}}$ obtained for some integer values of $\nu$ are presented in Table \ref{table2}. The direct comparison test determines the convergence of the series (see App. II).    

\begin{widetext}
\begin{center}
\begin{table}
\caption{List of the series $\sum^{\infty}_{n=1}\frac{\big[J_{\nu+1}(n\pi)\big]^{2}}{n^{2\nu}}=\frac{\pi^{2\nu-\frac{1}{2}}}{2^{2\nu+1}}\frac{\Gamma(2\nu)}{\Gamma\left(2\nu+\frac{3}{2}\right)\Gamma^{2}\left(\nu+\frac{1}{2}\right)}$.}
\begin{threeparttable}
 \begin{tabular}{|c|c|c| } 

 \hline\hline
  $\nu$ & $\varphi(x)$ & sum of the infinite series \\ [0.5ex] 
 \hline\hline \xrowht[()]{10pt}
 $1$ & $\sqrt{\frac{15}{2a^{5}}}x(a^{2}-x^{2})^{\frac{1}{2}}$  &  $\frac{4}{15}$  \\  \hline \xrowht[()]{10pt}
 
 $2$  & $\sqrt{\frac{315}{16a^{9}}}x(a^{2}-x^{2})^{\frac{3}{2}}$ &  $\frac{32\pi^{2}}{2835}$   \\ \hline \xrowht[()]{10pt}

 $3$  & $\sqrt{\frac{9009}{256a^{13}}}x(a^{2}-x^{2})^{\frac{5}{2}}$ &  $\frac{512\pi^{4}}{2027025}$   \\ \hline \xrowht[()]{10pt}

 $4$  & $\sqrt{\frac{109395}{2048a^{17}}}x(a^{2}-x^{2})^{\frac{7}{2}}$ &  $\frac{4096\pi^{6}}{1206079875}$   \\ \hline \xrowht[()]{10pt}

$5$  & $\sqrt{\frac{4849845}{65536a^{21}}}x(a^{2}-x^{2})^{\frac{9}{2}}$ & $\frac{131072\pi^{8}}{4331032831125}$   \\ \hline \xrowht[()]{10pt}

 $6$  & $\sqrt{\frac{50702925}{524288a^{25}}}x(a^{2}-x^{2})^{\frac{11}{2}}$ &  $\frac{1048576\pi^{10}}{5478756531373125}$   \\ \hline \xrowht[()]{10pt}

  $7$  & $\sqrt{\frac{1017958725}{8388608a^{29}}}x(a^{2}-x^{2})^{\frac{13}{2}}$ &  $\frac{16777216\pi^{12}}{18589420910949013125}$   \\ \hline \xrowht[()]{10pt}

 $8$  & $\sqrt{\frac{9917826435}{67108864a^{33}}}x(a^{2}-x^{2})^{\frac{15}{2}}$ &  $\frac{134217728\pi^{14}}{40750666268358943771875}$   \\ \hline \xrowht[()]{10pt}

   $9$  & $\sqrt{\frac{755505013725}{4294967296a^{37}}}x(a^{2}-x^{2})^{\frac{17}{2}}$ & $\frac{8589934592\pi^{16}}{897125917897922147137828125}$   \\ \hline \xrowht[()]{10pt}

   $10$  & $\sqrt{\frac{7064634602025}{34359738368a^{41}}}x(a^{2}-x^{2})^{\frac{19}{2}}$ &  $\frac{68719476736\pi^{18}}{3028398056850752528021595140625}$   \\ \hline

 \hline\hline 
\end{tabular}

\end{threeparttable}
\label{table2}
\end{table}
\end{center}
\end{widetext}

\begin{figure}
    \centering
    \includegraphics[width=0.8\linewidth]{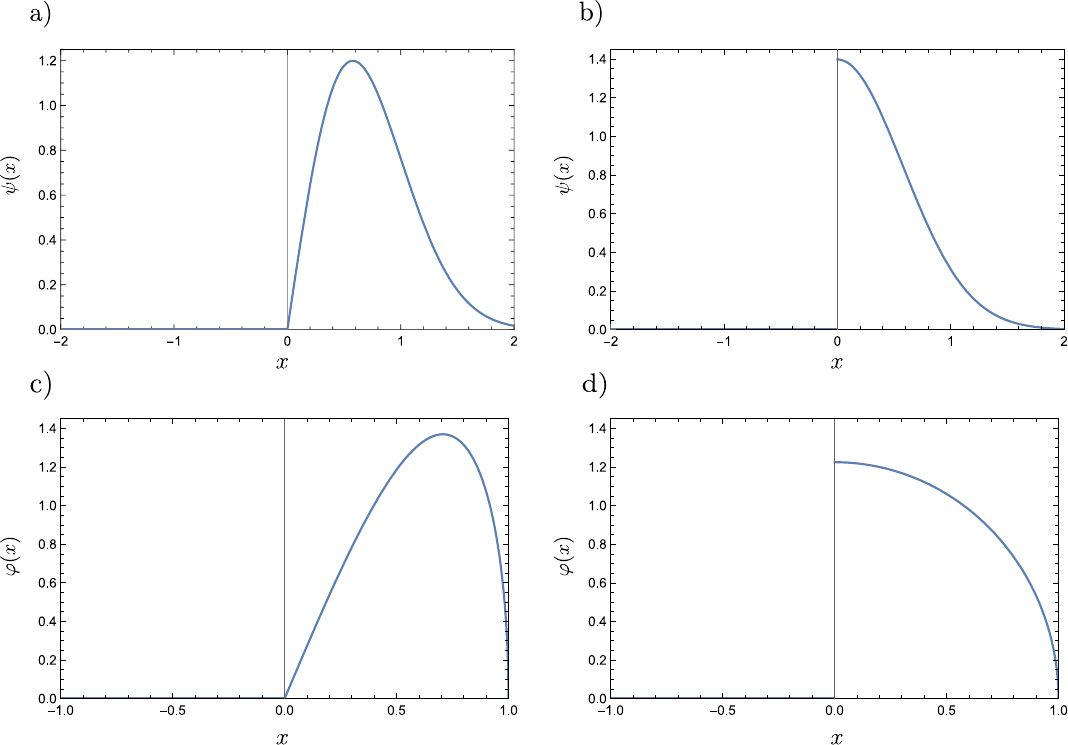}
    \caption{Wave functions \eqref{wf1} and \eqref{wf3} are shown in a) and c), respectively. Functions with discontinuity point $x=0$, \eqref{funkcgr} and \eqref{wf4}, are shown in b) and d). Functions \eqref{wf3} and \eqref{wf4} are plotted for $a=1$. Note that the functions shown in b) and d) do not represent regular wave functions since they are not continuous, but do, on the other hand, allow evaluation of additional infinite sums.} 
    \label{fig:1}
\end{figure}

We conclude by considering the function of the form
\begin{equation}
\varphi(x)=\sqrt{\frac{2}{\sqrt{\pi}}\frac{1}{a^{4\nu-1}}\frac{\Gamma\left(2\nu+\frac{1}{2}\right)}{\Gamma(2\nu)}}(a^{2}-x^{2})^{\nu-\frac{1}{2}}, \hspace{2mm} x\in(0,a) \hspace{1cm} {\rm{and}} \hspace{1cm}  \varphi(x)=0, \hspace{2mm} x\notin(0,a).    
\end{equation}
Note that this function is discontinuous at the point $x=0$ and therefore it does not represent a regular wave function. However, it allows us to calculate infinitely many sums of the form $\sum_{n=1}^{\infty}\frac{\left[\mathsf{H}_{\nu}(n\pi)\right]^{2}}{n^{2\nu}}$. Thus, using \cite{GRRI,erdel}
\begin{equation}
\int_{0}^{u}\left(u^{2}-x^{2}\right)^{\nu-\frac{1}{2}}\sin(\tilde{a}x)\d x=\frac{\sqrt{\pi}}{2}\left(\frac{2u}{\tilde{a}}\right)^{\nu}\Gamma\left(\nu+\frac{1}{2}\right)\mathsf{H}_{\nu}(\tilde{a}u), \hspace{2mm}u>0, \hspace{2mm} \mathrm{Re}\hspace{1mm}\nu>-\frac{1}{2}\label{struve}
\end{equation}
the function \begin{equation}
\varphi(x)=\sqrt{\frac{3}{2a^{3}}}(a^{2}-x^{2})^{\frac{1}{2}}, \hspace{2mm} x\in(0,a) \hspace{1cm} {\rm{and}} \hspace{1cm}  \varphi(x)=0, \hspace{2mm} x\notin(0,a)   \label{wf4}
\end{equation}
obtained for $\nu=1$ leads to coefficients
\begin{equation}
C_{n}=\frac{\sqrt{3}}{2}\frac{\mathsf{H}_{1}(n\pi)}{n},    
\end{equation}
wherefrom we evaluate the infinite sum
\begin{equation}
\sum_{n=1}^{\infty}\frac{\left[\mathsf{H}_{1}(n\pi)\right]^{2}}{n^{2}}=\frac{4}{3}.    
\end{equation}
Functions \eqref{wf3} and \eqref{wf4} are shown in Figure \ref{fig:1} c) and d), respectively.
In Table \ref{table3} we list several examples of sums of the form $\sum_{n=1}^{\infty}\frac{\left[\mathrm{H}_{\nu}(n\pi)\right]^{2}}{n^{2\nu}}$. The convergence of these sums is additionally verified in App. II.

\begin{widetext}
\begin{center}
\begin{table}
\caption{List of the series $\sum_{n=1}^{\infty}\frac{\left[\mathrm{H}_{\nu}(n\pi)\right]^{2}}{n^{2\nu}}=\frac{\pi^{2\nu-\frac{1}{2}}}{2^{2\nu}}\frac{\Gamma(2\nu)}{\Gamma\left(2\nu+\frac{1}{2}\right)\Gamma^{2}\left(\nu+\frac{1}{2}\right)}$.}
\begin{threeparttable}
 \begin{tabular}{|c|c|c| }
 \hline\hline
  $\nu$ & $\varphi(x)$ & sum of the infinite series \\ [0.5ex] 
 \hline\hline \xrowht[()]{10pt}
 $1$ & $\sqrt{\frac{3}{2a^{3}}}(a^{2}-x^{2})^{\frac{1}{2}}$ &  $\frac{4}{3}$  \\  \hline \xrowht[()]{10pt}
 
 $2$  & $\sqrt{\frac{35}{16a^{7}}}(a^{2}-x^{2})^{\frac{3}{2}}$ &  $\frac{32\pi^{2}}{315}$   \\ \hline \xrowht[()]{10pt}

 $3$ & $\sqrt{\frac{693}{256a^{11}}}(a^{2}-x^{2})^{\frac{5}{2}}$ &  $\frac{512\pi^{4}}{155925}$   \\ \hline \xrowht[()]{10pt}

 $4$  & $\sqrt{\frac{6435}{2048a^{15}}}(a^{2}-x^{2})^{\frac{7}{2}}$ &  $\frac{4096\pi^{6}}{70945875}$   \\ \hline \xrowht[()]{10pt}

$5$  & $\sqrt{\frac{230945}{65536a^{19}}}(a^{2}-x^{2})^{\frac{9}{2}}$  & $\frac{131072\pi^{8}}{206239658625}$   \\ \hline \xrowht[()]{10pt}

 $6$  & $\sqrt{\frac{2028117}{524288a^{23}}}(a^{2}-x^{2})^{\frac{11}{2}}$ & $\frac{1048576\pi^{10}}{219150261254925}$   \\ \hline \xrowht[()]{10pt}

  $7$  & $\sqrt{\frac{35102025}{8388608a^{27}}}(a^{2}-x^{2})^{\frac{13}{2}}$ & $\frac{16777216\pi^{12}}{641014514170655625}$   \\ \hline \xrowht[()]{10pt}

 $8$  & $\sqrt{\frac{300540195}{67108864a^{31}}}(a^{2}-x^{2})^{\frac{15}{2}}$ &  $\frac{134217728\pi^{14}}{1234868674798755871875}$   \\ \hline \xrowht[()]{10pt}

   $9$  & $\sqrt{\frac{20419054425}{4294967296a^{35}}}(a^{2}-x^{2})^{\frac{17}{2}}$ &  $\frac{8589934592\pi^{16}}{24246646429673571544265625}$   \\ \hline \xrowht[()]{10pt}

   $10$  & $\sqrt{\frac{172308161025}{34359738368a^{39}}}(a^{2}-x^{2})^{\frac{19}{2}}$ & $\frac{68719476736\pi^{18}}{73863367240262256781014515625}$   \\ \hline

 \hline\hline 
\end{tabular}

\end{threeparttable}
\label{table3}
\end{table}
\end{center}
\end{widetext}

For more details on the calculation of integrals \eqref{besel} and \eqref{struve}, see App. I.

\section{Conclusion}

In the present paper we have calculated various previously unevaluated nontrivial infinite sums analytically by the application of the basic quantum mechanical principles to simple physical models of a half harmonic oscillator and a particle trapped inside an infinite potential well. For that purpose, we used well-defined wave functions to evaluate sums of the type: $\sum^{\infty}_{n=0}\frac{2^{2n}}{(2n+1)!}\Gamma^{2}\left(n+\frac{3}{2}\right)\left[\hspace{0.2mm}_2\hspace{-0.03cm}F_1\left(-n,\frac{\nu+2}{2};\frac{3}{2};\frac{1}{2}\right)\right]^{2}$, $\sum^{\infty}_{n=0}\frac{\left[L_{\nu}^{2n+1-\nu}\left(\frac{b^{2}}{2}\right)\right]^{2}b^{4n}}{2^{2n}(2n+1)!}$ and $\sum^{\infty}_{n=1}\frac{\big[J_{\nu+1}(n\pi)\big]^{2}}{n^{2\nu}}$, where $\nu$ is an integer. Some examples of these sums of infinite series are presented in Tables \ref{table1}, \ref{table4} and \ref{table2}. Also, we were able to calculate infinite series $\sum^{\infty}_{n=0}\frac{2^{2n}}{(2n+1)!}\Gamma^{2}\left(n+\frac{3}{2}\right)\left[\hspace{0.5mm}_2\hspace{-0.05cm}F_1\left(-n,1;\frac{3}{2};\frac{1}{2}\right)\right]^{2}$ (see \eqref{diskon}) as well as sums of the type $\sum_{n=1}^{\infty}\frac{\left[\mathsf{H}_{\nu}(n\pi)\right]^{2}}{n^{2\nu}}$, some of which are presented in Table \ref{table3}, starting from the functions with a discontinuity point at one of the interval boundaries as one point does not disrupt the convergence of series secured by the properties of the corresponding Hilbert space. This leaves room for using analogous procedure to calculate additional sums starting from functions which are, strictly speaking, not regular wave functions aside from using well-defined quantum mechanical problems. 
Let us emphasize that though the convergence of the evaluated infinite sums has been additionally verified by different convergence tests, the main advantage of the method used throughout this paper is that the convergence of these sums is secured even if they cannot be calculated analytically or their convergence is difficult to be proven by convergence tests.
\section*{Appendix I} \label{app1}
In order to prove identity \eqref{hipergeo}, we start from the integral representation of the Gaussian hypergeometric function in terms of Kummer's hypergeometric function \cite{GRRI}:
\begin{equation}
    _2F_1(a,b;c;z)=\frac{1}{\Gamma(b)}\int_0^{\infty}\mbox{e}^{-t}t^{b-1}\hspace{0,1cm}_1F_1(a;c;tz)\d t\,.
\end{equation}
Hence, we obtain
\begin{equation}
    _2F_1\left(-n,\frac{\nu}{2}+1;\frac 32;\frac 12\right)=\frac{1}{\Gamma\left(\frac{\nu}{2}+1\right)}\int_0^{\infty}\mbox{e}^{-t}t^{\frac{\nu}{2}}\hspace{0.1cm}_1F_1\left(-n;\frac 32;\frac t2\right)\d t\,.
\end{equation}
Using substitution $t=2x^2$, we get

\begin{equation}
    _2F_1\left(-n,\frac{\nu}{2}+1;\frac 32;\frac 12\right)\Gamma\left(\frac{\nu}{2}+1\right)=\int_0^{\infty}\mbox{e}^{-2x^2}2^{\frac{\nu}{2}+2}x^{\nu+1}\hspace{0.1cm}_1F_1\left(-n;\frac 32;x^2\right)\d x\,.
\end{equation}
We now use the relation between Hermite polynomial and Kummer's hypergeometric function \cite{GRRI}:
\begin{equation}
   H_{2n+1}(x)=(-1)^n \frac{(2n+1)!}{n!}2x \hspace{0.1cm}_1F_1\left(-n;\frac 32;x^2\right)
\end{equation}
wherefrom we obtain

\begin{equation}
    _2F_1\left(-n,\frac{\nu}{2}+1;\frac 32;\frac 12\right)\Gamma\left(\frac{\nu}{2}+1\right)=\int_0^{\infty}\mbox{e}^{-2x^2}2^{\frac{\nu}{2}+1}x^{\nu}\frac{\mbox{H}_{2n+1}(x)n!}{(-1)^n(2n+1)!}\d x\,.
\end{equation}
Finally, using the expression for the gamma function for half-integer values
\begin{equation}
    \Gamma\left(\frac 12 +n\right)=\frac{(2n)!}{2^{2n}n!}\sqrt{\pi}
\end{equation}
we obtain identity \eqref{hipergeo}.

The identities \eqref{sinus} and \eqref{kosinus} will be proven by using series expansions of sine and cosine functions \cite{zbirka,cos}:
\begin{equation}
    \sin(bx)=\e^{-\frac{b^2}{4}}\sum_{n=0}^{\infty}\frac{(-1)^n b^{2n+1}}{2^{2n+1}(2n+1)!}H_{2n+1}(x),
\end{equation}
\begin{equation}
    \cos(bx)=\e^{-\frac{b^2}{4}}\sum_{n=0}^{\infty}\frac{(-1)^n b^{2n}}{2^{2n}(2n)!}H_{2n}(x)\,.
\end{equation}
Starting from
\begin{equation}
\int_0^{\infty}\e^{-x^2}H_{2n+1}(x)\sin(bx)\d x=\frac 12 \e^{-\frac{b^2}{4}}\sum_{m=0}^{\infty}\frac{(-1)^m b^{2m+1}}{2^{2m+1}(2m+1)!}\int_{-\infty}^{\infty}\e^{-x^2}H_{2n+1}(x)H_{2m+1}(x)\d x\,,
\end{equation}
by employing the orthogonality condition  
\begin{equation}
    \int_{-\infty}^{\infty}\e^{-x^2}H_n(x)H_m(x)\d x
=2^n n! \sqrt{\pi}\delta_{n,m}\end{equation}
we get equation \eqref{sinus}. Analogously, for expression
\begin{equation}
\int_{0}^{\infty} \e^{-x^2}H_{2n+1}(x) x \cos(bx)\d x=\frac 12 \e^{-\frac{b^2}{4}}\sum_{m=0}^{\infty}\frac{(-1)^m b^{2m}}{2^{2m}(2m)!}\int_{-\infty}^{\infty}x \e^{-x^2}H_{2n+1}(x)H_{2m}(x)\d x
\end{equation}
by making use of \cite{zbirka}
\begin{equation}
    \int_{0}^{\infty}x\e^{-x^2}H_n(x) H_m(x)\d x=2^{n-1}n!\sqrt{\pi}\big(\delta_{n-1,m}+2(n+1)\delta_{n+1,m}\big)
\end{equation}
we get equation \eqref{kosinus}. Further on, successive use of recurrence relation
\begin{equation}
xH_{n}(x)=\frac{1}{2}H_{n+1}(x)+nH_{n-1}(x)    
\end{equation}
leads to the expressions:
\begin{equation}
\int_{0}^{\infty}\e^{-x^{2}}H_{2n+1}(x)H_{2m}(x)\sin(bx)\d x=\frac{(-1)^{n+m}}{2}(2m)!2^{2m}\sqrt{\pi}b^{2n+1-2m}\e^{-\frac{b^{2}}{4}}L_{2m}^{2n+1-2m}\left(\frac{b^{2}}{2}\right)   \label{sin} 
\end{equation}
and
\begin{equation}
\int_{0}^{\infty}\e^{-x^{2}}H_{2n+1}(x)H_{2m+1}(x)\cos(bx)\d x=\frac{(-1)^{n+m}}{2}(2m+1)!2^{2m+1}\sqrt{\pi}b^{2n-2m}\e^{-\frac{b^{2}}{4}}L_{2m+1}^{2n-2m}\left(\frac{b^{2}}{2}\right) \label{cos}   
\end{equation}
used when calculating the sums of infinite series presented in Table \ref{table4}.

We proceed to prove the identity \eqref{besel}. Therefore, we start with the integral representation of the Bessel function \cite{GRRI}:
\begin{equation}
 J_{\nu}(z)=\frac{1}{\Gamma(\nu+\frac{1}{2})\sqrt{\pi}}\left(\frac{z}{2}\right)^{\nu}\int^1_{-1}(1-t^2)^{\nu-\frac{1}{2}}\e^{\mathrm{i} zt}\d t, \quad \mathrm{Re}\,\nu>-\frac{1}{2}\,.
\end{equation}
By putting $\nu\to \nu+1$ and $z=\tilde{a}u$ we obtain
\begin{equation}
 J_{\nu+1}(\tilde{a}u)=\frac{1}{\Gamma(\nu+\frac{3}{2})\sqrt{\pi}}\left(\frac{\tilde{a}u}{2}\right)^{\nu+1}\int^1_{-1}(1-t^2)^{\nu+\frac{1}{2}}\e^{\mathrm{i} \tilde{a}ut}\d t, \quad \mathrm{Re}\,\nu>-\frac{1}{2}\,.
\end{equation}
Using identity $\Gamma\left(\nu+\frac 32\right)=\left(\nu+\frac 12\right)\Gamma\left(\nu+\frac 12\right)$ and substitution $t=\frac lu$ we arrive at 
\begin{equation}
    J_{\nu+1}(\tilde{a}u)\left(\nu+\frac 12\right)\Gamma\left(\nu+\frac 12\right)\sqrt{\pi}\left(\frac{2u}{\tilde{a}}\right)^{\nu+1}=2\int_0^u
(u^2-l^2)^{\nu+\frac 12}\cos(\tilde{a}l)\d l\,.\end{equation}
Performing partial integration we obtain the identity \eqref{besel}.

We finally prove \eqref{struve}. To that end, we
start from the integral representation of the Struve function of the first kind \cite{GRRI}:
\begin{equation}
    \mathsf{H}_{\nu}(x)=\frac{2(\frac{x}{2})^{\nu}}{\sqrt{\pi}\Gamma(\nu+\frac 12)}\int_0^1(1-t^2)^{\nu-\frac 12}\sin{(xt)}\d t, \quad \mbox{Re}\,\nu>-\frac 12,
\end{equation}
wherefrom by putting $x=\tilde{a}u$ we get
\begin{equation}
    \mathsf{H}_{\nu}(\tilde{a}u)\frac{\sqrt{\pi}}{2}\Gamma\left(\nu+\frac 12\right)\left(\frac{2}{\tilde{a}u}\right)^{\nu}=\int_0^1(1-t^2)^{\nu-\frac 12}\sin{(\tilde{a}ut)}\d t\,.
\end{equation}    
Using substitution $ut=x$, we easily obtain
\begin{equation}
    \mathsf{H}_{\nu}(\tilde{a}u)\frac{\sqrt{\pi}}{2}\Gamma\left(\nu+\frac 12\right)\left(\frac{2u}{\tilde{a}}\right)^{\nu}=\int_0^u(u^2-x^2)^{\nu-\frac 12}\sin{(\tilde{a}x)}\d x\,.
\end{equation}

\section*{Appendix II} \label{app2}

In order to show the convergence of the series $\sum^{\infty}_{n=0}\frac{2^{2n}}{(2n+1)!}\Gamma^{2}\left(n+\frac{3}{2}\right)\left[\hspace{0.2mm}_2\hspace{-0.03cm}F_1\left(-n,\frac{3}{2};\frac{3}{2};\frac{1}{2}\right)\right]^{2}=\sum_{n=0}^{\infty}a_{n}$, one can use d'Alembert's ratio test
\begin{equation}
\lim_{n\to\infty}\left|\frac{a_{n+1}}{a_{n}}\right|<1, \label{dlmb}   
\end{equation}
where $a_{n}$ is the $n$th term of the discussed sum. This leads to
\begin{equation}
\lim_{n\to\infty}\left|\frac{a_{n+1}}{a_{n}}\right|=\lim_{n\to\infty}\left[\frac{\frac{2^{2n+2}}{(2n+3)!}\Gamma^{2}\left(n+\frac{5}{2}\right)}{\frac{2^{2n}}{(2n+1)!}\Gamma^{2}\left(n+\frac{3}{2}\right)}\frac{\left[\hspace{0.2mm}_2\hspace{-0.03cm}F_1\left(-n-1,\frac{3}{2};\frac{3}{2};\frac{1}{2}\right)\right]^{2}}{\left[\hspace{0.2mm}_2\hspace{-0.03cm}F_1\left(-n,\frac{3}{2};\frac{3}{2};\frac{1}{2}\right)\right]^{2}}\right].    
\end{equation}
Since $\Gamma\left(n+\frac{5}{2}\right)=\left(n+\frac{3}{2}\right)\Gamma\left(n+\frac{3}{2}\right)$, 
one arrives at
\begin{equation}
\frac{\frac{2^{2(n+1)}}{(2n+3)!}}{\frac{2^{2n}}{(2n+1)!}}\left[\frac{\Gamma\left(n+\frac{5}{2}\right)}{\Gamma\left(n+\frac{3}{2}\right)}\right]^{2}=\frac{4}{(2n+3)(2n+2)}\left[\frac{\left(n+\frac{3}{2}\right)\Gamma\left(n+\frac{3}{2}\right)}{\Gamma\left(n+\frac{3}{2}\right)}\right]^{2}=\frac{4\left(n+\frac{3}{2}\right)^{2}}{(2n+3)(2n+2)}.   
\end{equation}
Also, using \eqref{hyper3/2} one can write
\begin{equation}
\lim_{n\to\infty}\left|\frac{a_{n+1}}{a_{n}}\right|=\lim_{n\to\infty}\left[\frac{4\left(n+\frac{3}{2}\right)^{2}}{(2n+3)(2n+2)}\left(\frac{\left(\frac{1}{2}\right)^{n+1}}{\left(\frac{1}{2}\right)^{n}}\right)^{2}\right]=\frac{1}{4}<1,    
\end{equation}
This proves the convergence of the series $\sum^{\infty}_{n=0}\frac{2^{2n}}{(2n+1)!}\Gamma^{2}\left(n+\frac{3}{2}\right)\left[\hspace{0.2mm}_2\hspace{-0.03cm}F_1\left(-n,\frac{3}{2};\frac{3}{2};\frac{1}{2}\right)\right]^{2}$. By applying analogous procedure, one can prove the convergence of the other sums of type $\sum^{\infty}_{n=0}\frac{2^{2n}}{(2n+1)!}\Gamma^{2}\left(n+\frac{3}{2}\right)\left[\hspace{0.2mm}_2\hspace{-0.03cm}F_1\left(-n,\frac{\nu+2}{2};\frac{3}{2};\frac{1}{2}\right)\right]^{2}$ for odd values of $\nu$ from Table \ref{table1}. Since d'Alembert's convergence test turns out to be inconclusive for even values of $\nu$, the convergence of this type of infinite series can be verified by Raabe–Duhamel's test. Since $\lbrace\frac{2^{2n}}{(2n+1)!}\Gamma^{2}\left(n+\frac{3}{2}\right)\left[\hspace{0.2mm}_2\hspace{-0.03cm}F_1\left(-n,\alpha;\beta;z\right)\right]^{2}\rbrace$ is a sequence of positive numbers, convergence of the discussed series is confirmed if the following condition is satisfied
\begin{equation}
\lim_{n\to\infty}\left[n\left(\frac{b_{n}}{b_{n+1}}-1\right)\right]>1.    
\end{equation}
We start by writing down $\sum^{\infty}_{n=0}\frac{2^{2n}}{(2n+1)!}\Gamma^{2}(n+\frac{3}{2})\left[\hspace{0.2mm}_2\hspace{-0.03cm}F_1\left(-n,\alpha;\beta;z\right)\right]^{2}$ in the following form:
\begin{equation}
\sum^{\infty}_{n=0}\frac{2^{2n}}{(2n+1)!}\Gamma^{2}\left(n+\frac{3}{2}\right)\left[\hspace{0.2mm}_2\hspace{-0.03cm}F_1\left(-n,\alpha;\beta;z\right)\right]^{2}=\sum_{n=0}^{\infty}b_{n}=\sum_{n=0}^{\infty}\tilde{a}_{n}\tilde{b}_{n},   
\end{equation}
where 
\begin{equation}
\tilde{a}_{n}=\frac{2^{2n}}{(2n+1)!}\Gamma^{2}\left(n+\frac{3}{2}\right), \hspace{2mm} \tilde{b}_{n}=\left[\hspace{0.2mm}_2\hspace{-0.03cm}F_1\left(-n,\alpha;\beta;z\right)\right]^{2}.    
\end{equation}
Raabe–Duhamel's test now takes the form
\begin{equation}
\lim_{n\to\infty}\left[n\left(\frac{\tilde{a}_{n}\tilde{b}_{n}}{\tilde{a}_{n+1}\tilde{b}_{n+1}}-1\right)\right]>1.    \label{rdt}
\end{equation}
For $n\to\infty$ ratio $\frac{\tilde{a}_{n}}{\tilde{a}_{n+1}}$ reduces to
\begin{align}
\frac{\tilde{a}_{n}}{\tilde{a}_{n+1}}=\frac{\frac{2^{2n}}{(2n+1)!}}{\frac{2^{2(n+1)}}{(2n+3)!}}\left[\frac{\Gamma\left(n+\frac{3}{2}\right)}{\Gamma\left(n+\frac{5}{2}\right)}\right]^{2}=\frac{(2n+3)(2n+2)}{4\left(n+\frac{3}{2}\right)^{2}}=1-\frac{1}{2n}+\mathcal{O}\left(\frac{1}{n^{2}}\right). \label{ratioa}    
\end{align}
The series expansion of hypergeometric function $\hspace{0.2mm}_2\hspace{-0.03cm}F_1\left(-n-1,\alpha;\beta;z\right)$ reads
\begin{align}
\hspace{0.2mm}_2\hspace{-0.03cm}F_1\left(-n-1,\alpha;\beta;z\right)&=\sum_{l=0}^{\infty}\frac{(-n-1)_{l}(\alpha)_{l}\, z^{l}}{\left(\beta\right)_{l}\,l!}=\sum_{l=0}^{\infty}\frac{(n+1)!(-1)^{l}(\alpha)_{l}\, z^{l}}{(n+1-l)!\left(\beta\right)_{l}\,l!}=\sum_{l=0}^{\infty}\frac{(-n)_{l}(\alpha)_{l}\,z^{l}}{(\beta)_{l}\,l!}\frac{1}{1-\frac{l}{n+1}},
\end{align}
where we used $(-n)_{l}=\frac{n!(-1)^{l}}{(n-l)!}$. We will keep the leading terms in the following expansion:
\begin{equation}
\frac{1}{1-\frac{l}{n+1}}=1+\frac{l}{n+1}+\mathcal{O}\left(\frac{l^{p}}{n^{q}}\right), \hspace{2mm} p=1,2,3,... \hspace{2mm} q=2,3,4,...
\end{equation}
Using substitution $k=l-1$ we obtain
\begin{align}
\hspace{0.2mm}_2\hspace{-0.03cm}F_1\left(-n-1,\alpha;\beta;z\right)&=\hspace{0.2mm}_2\hspace{-0.03cm}F_1\left(-n,\alpha;\beta;z\right)+\frac{1}{n+1}\sum_{k=0}^{\infty}\frac{(-n)_{k+1}(\alpha)_{k+1}\, z^{k+1}}{(\beta)_{k+1}\,k!}+\mathcal{O}\left(\frac{1}{n^{2}}\right)\nonumber \\ &=\hspace{0.2mm}_2\hspace{-0.03cm}F_1\left(-n,\alpha;\beta;z\right)-\left(1-\frac{1}{n}\right)\frac{\alpha z}{\beta}\hspace{0.2mm}_2\hspace{-0.03cm}F_1\left(-n+1,\alpha+1;\beta+1;z\right)+\mathcal{O}\left(\frac{1}{n^{2}}\right),   \label{hyper}
\end{align}
where we utilized 
\begin{equation}
(\alpha)_{k+1}=\alpha(\alpha+1)_{k}, \hspace{2mm} (\beta)_{k+1}=\beta(\beta+1)_{k}, \hspace{2mm} (-n)_{k+1}=-n(-n+1)_{k}.
\end{equation}
For $n\to\infty$ the relation \eqref{hyper} leads to
\begin{equation}
\frac{\hspace{0.2mm}_2\hspace{-0.03cm}F_1\left(-n,\alpha;\beta;z\right)}{\hspace{0.2mm}_2\hspace{-0.03cm}F_1\left(-n-1,\alpha;\beta;z\right)}=1+\left(1-\frac{1}{n}\right)\frac{\alpha z}{\beta}\frac{\hspace{0.2mm}_2\hspace{-0.03cm}F_1\left(-n,\alpha+1;\beta+1;z\right)}{\hspace{0.2mm}_2\hspace{-0.03cm}F_1\left(-n,\alpha;\beta;z\right)}+\mathcal{O}\left(\frac{1}{n^{2}}\right).  \label{hyper2}   
\end{equation}
The ratio $\frac{\hspace{0.2mm}_2\hspace{-0.03cm}F_1\left(-n,\alpha+1;\beta+1;z\right)}{\hspace{0.2mm}_2\hspace{-0.03cm}F_1\left(-n,\alpha;\beta;z\right)}$ for $n\to\infty$ is given by \cite{cuyt}
\begin{equation}
\frac{\hspace{0.2mm}_2\hspace{-0.03cm}F_1\left(-n,\alpha+1;\beta+1;z\right)}{\hspace{0.2mm}_2\hspace{-0.03cm}F_1\left(-n,\alpha;\beta;z\right)}\sim\frac{\beta}{nz},   
\end{equation}
wherefrom \eqref{hyper2} becomes
\begin{equation}
\frac{\hspace{0.2mm}_2\hspace{-0.03cm}F_1\left(-n,\alpha;\beta;z\right)}{\hspace{0.2mm}_2\hspace{-0.03cm}F_1\left(-n-1,\alpha;\beta;z\right)}=1+\frac{\alpha}{n}+\mathcal{O}\left(\frac{1}{n^{2}}\right)   
\end{equation}
leading to
\begin{equation}
\frac{\tilde{b}_{n}}{\tilde{b}_{n+1}}=\left[\frac{\hspace{0.2mm}_2\hspace{-0.03cm}F_1\left(-n,\alpha;\beta;z\right)}{\hspace{0.2mm}_2\hspace{-0.03cm}F_1\left(-n-1,\alpha;\beta;z\right)}\right]^{2}=1+\frac{2\alpha}{n}+\mathcal{O}\left(\frac{1}{n^{2}}\right). \label{ratiob}    
\end{equation}
Finally, taking into account \eqref{ratioa} and \eqref{ratiob}, the convergence criterion \eqref{rdt} reduces to
\begin{equation}
\lim_{n\to\infty}\left[n\left(\frac{b_{n}}{b_{n+1}}-1\right)\right]=\frac{4\alpha-1}{2}>1,    
\end{equation}
which is satisfied for $\alpha>\frac{3}{4}$. For example, in case $\alpha=1$, $\beta=\frac{3}{2}$ and $z=\frac{1}{2}$ (the second sum in Table \ref{table1}) the previous expression becomes
\begin{equation}
\lim_{n\to\infty}\left[n\left(\frac{b_{n}}{b_{n+1}}-1\right)\right]=\frac{3}{2}>1. 
\end{equation}

We proceed to prove the convergence of the sum $\sum_{n=0}^{\infty}\frac{\left[L_{0}^{2n+1}\left(\frac{b^{2}}{2}\right)\right]^{2}b^{4n}}{2^{2n}(2n+1)!}=\sum_{n=1}^{\infty}d_{n}$. Since $L_{0}^{2n+1}\left(\frac{b^{2}}{2}\right)=1$ for all $n$, by using d'Alembert's ratio test \eqref{dlmb}, it is straightforward to show the convergence of this sum:
\begin{equation}
\lim_{n\to\infty}\left|\frac{d_{n+1}}{d_{n}}\right|=\lim_{n\to\infty}\frac{\frac{b^{4n+4}}{2^{2n+2}(2n+3)!}}{\frac{b^{4n}}{2^{2n}(2n+1)!}}=0.   
\end{equation}
The next infinite sum from Table \ref{table4} reads $\sum_{n=1}^{\infty}\frac{\left[L_{1}^{2n}\left(\frac{b^{2}}{2}\right)\right]^{2}b^{4n}}{2^{2n}(2n+1)!}=\sum_{n=0}^{\infty}f_{n}$. By expressing the associated Laguerre polynomials as \cite{hand}
\begin{equation}
L_{p}^{q}(z)=\frac{\Gamma(p+q+1)}{\Gamma(q+1)p!}\hspace{0.2mm}_1\hspace{-0.03cm}F_1\left(-p;q+1;z\right),    
\end{equation}
the ratio test leads to
\begin{equation}
\lim_{n\to\infty}\left|\frac{f_{n+1}}{f_{n}}\right|=\lim_{n\to\infty}\left[\left(\frac{\frac{\Gamma(2n+4)}{\Gamma(2n+3)}}{\frac{\Gamma(2n+2)}{\Gamma(2n+1)}}\right)^{2}\left(\frac{\hspace{0.2mm}_1\hspace{-0.03cm}F_1\left(-1;2n+3;\frac{b^{2}}{2}\right)}{\hspace{0.2mm}_1\hspace{-0.03cm}F_1\left(-1;2n+1;\frac{b^{2}}{2}\right)}\right)^{2}\frac{\frac{b^{4n+4}}{2^{2n+2}(2n+3)!}}{\frac{b^{4n}}{2^{2n}(2n+1)!}}\right]=0,   
\end{equation}
where we used $\Gamma(n+1)=n!$ for $n\in\mathbb{N}_0$ and
\begin{equation}
\hspace{0.2mm}_1\hspace{-0.03cm}F_1\left(-1;2n+3;\frac{b^{2}}{2}\right)=1-\frac{\frac{b^{2}}{2}}{2n+3}, \hspace{2mm} \hspace{0.2mm}_1\hspace{-0.03cm}F_1\left(-1;2n+1;\frac{b^{2}}{2}\right)=1-\frac{\frac{b^{2}}{2}}{2n+1}.    
\end{equation}
The convergence of the succeeding sums in Table \ref{table4} can be verified analogously.

Let us now prove the convergence of the infinite sums of type $\sum^{\infty}_{n=1}\frac{\big[J_{\nu+1}(n\pi)\big]^{2}}{n^{2\nu}}$ by exploiting the direct comparison test. Namely, since $|J_{\nu}(z)|\leq 1$ for all $z$ \cite{landau}, one can write
\begin{equation}
\frac{\left[J_{\nu+1}(n\pi)\right]^{2}}{n^{2\nu}}\leq\frac{1}{n^{2}}, \hspace{2mm} \nu=1,2,3,...
\end{equation}
Taking into account that the series $\sum_{n=1}^{\infty}\frac{1}{n^{2}}=\frac{\pi^{2}}{6}$ is absolutely convergent \cite{euler},
 the series $\sum^{\infty}_{n=1}\frac{\big[J_{\nu+1}(n\pi)\big]^{2}}{n^{2\nu}}$ is also convergent.

Finally, the convergence of the infinite sums of type $\sum^{\infty}_{n=1}\frac{\big[\mathsf{H}_{\nu}(n\pi)\big]^{2}}{n^{2\nu}}$ is proven by direct comparison test. By plotting functions $\frac{\big[\mathsf{H}_{\nu}(n\pi)\big]^{2}}{n^{2\nu}}$ and $\frac{2}{n^{2}}$ one can easily see that
\begin{equation}
\frac{\big[\mathsf{H}_{\nu}(n\pi)\big]^{2}}{n^{2\nu}}\leq\frac{2}{n^{2}},  \hspace{2mm} \nu=1,2,3,...    
\end{equation}
and bearing in mind that the series $\sum_{n=1}^{\infty}\frac{1}{n^{2}}=\frac{\pi^{2}}{6}$ is absolutely convergent, the discussed series involving Struve functions of the first kind is convergent as well.

\begin{acknowledgments}
The authors gratefully acknowledge the financial support of the Ministry of Science, Technological Development and Innovation of the Republic of Serbia (Grants No. 451-03-137/2025-03/200125 and 451-03-136/2025-03/200125).
\end{acknowledgments}


\end{document}